\documentclass[aps,amsfonts,amssymb,nofootinbib,twocolumn,amsmath,superscriptaddress]{revtex4-2}
\usepackage{graphicx}
\usepackage{bm}
\usepackage{hyperref}
\usepackage{IEEEtrantools}
\usepackage{natbib}
\usepackage{float}
\usepackage{siunitx}
\usepackage{xcolor}
\hypersetup{
    colorlinks,
    linkcolor={red!50!black},
    citecolor={blue!50!black},
    urlcolor={blue!80!black}
}
\makeatletter
\newcommand*{\balancecolsandclearpage}{%
  \close@column@grid
  \cleardoublepage
  \twocolumngrid
}
\makeatother
\restylefloat{table}

\begin{document}

\title{Anomalous high-temperature THz nonlinearity in superconductors near the metal-insulator transition}
 
\author{Dipanjan Chaudhuri}\email{dc36@illinois.edu}
\affiliation{Department of Physics and Astronomy, The Johns Hopkins University, Baltimore, Maryland 21218, USA}

\author{David Barbalas}
\affiliation{Department of Physics and Astronomy, The Johns Hopkins University, Baltimore, Maryland 21218, USA}

\author{Ralph Romero III}
\affiliation{Department of Physics and Astronomy, The Johns Hopkins University, Baltimore, Maryland 21218, USA}

\author{Fahad Mahmood}
\affiliation{Department of Physics and Astronomy, The Johns Hopkins University, Baltimore, Maryland 21218, USA}
\affiliation{Department of Physics, University of Illinois at Urbana-Champaign, Urbana, 61801 IL, USA}
\affiliation{F. Seitz Materials Research Laboratory, University of Illinois at Urbana-Champaign, Urbana, 61801 IL, USA}

\author{Jiahao Liang}
\affiliation{Department of Physics and Astronomy, The Johns Hopkins University, Baltimore, Maryland 21218, USA}

\author{John Jesudasan}
\affiliation{Department of Condensed Matter and Material Science, Tata Institute of Fundamental Research \\ 1, Homi Bhabha Rd., Colaba, Mumbai 400005, India}

\author{Pratap Raychaudhuri}
\affiliation{Department of Condensed Matter and Material Science, Tata Institute of Fundamental Research \\ 1, Homi Bhabha Rd., Colaba, Mumbai 400005, India}

\author{N. P. Armitage}\email{npa@jhu.edu}
\affiliation{Department of Physics and Astronomy, The Johns Hopkins University, Baltimore, Maryland 21218, USA}

\date{\today}

\begin{abstract}
The interplay of strong disorder and superconductivity is a topic of long-term interest in condensed matter physics.  Here we explore the nonlinear THz response of superconducting NbN films close to the 3D metal-insulator transition.  For the least disordered samples, the magnitude of the nonlinear $\chi^{(3)}$ response follows the temperature dependence of the superfluid density as expected.   In contrast, for high disorder samples near the metal-insulator transition the $\chi^{(3)}$ nonlinearity persists to temperatures as high as even 4 times the $T_c$ of the cleanest sample.   We discuss the possible origins of this remarkably large nonlinearity, including the possibility that it arises in an enhancement of the temperature scales of superconductivity close to localization.  Our work highlights the importance of finite frequency nonlinear THz experiments in detecting superconducting correlations even into regions where long-range ordered superconductivity does not persist.
\end{abstract}

\maketitle

The effect of disorder on superconductivity is a fascinating problem that lies at the intersection of several fundamentally important phenomena in many-body electronic systems. Attractive interactions of electrons close to the Fermi surface leads to superconductivity and is nearly ubiquitous in metals at low temperatures \cite{bardeen1957theory,de2018superconductivity}. Localization is also a canonical property of disordered electronic systems, which leads to an insulating ground state at low temperatures \cite{anderson1958absence}. The mutual interplay of these opposing phenomena remains a problem of significant interest from both theoretical and experimental perspective.

Anderson first addressed the problem of nonmagnetic impurities in a superconductor where he posited that the BCS framework can be generalized to a theory of disordered superconductors where Cooper pairing of $(k,\uparrow)$ and $(k,\downarrow)$ electrons is replaced by pairing of the exact time-reversed partners of one-electron eigenstates in disordered systems when $k$ is not a good quantum number \cite{anderson1959theory}. Abrikosov and Gor'kov independently arrived at the same conclusion as  \cite{abrikosov1959theory, abrikosov1959superconducting} ``Anderson's theorem" that the crossover from free electronic motion in a clean system to a diffusive one in presence of disorder does not appreciably change the critical temperature ($T_c$) of the superconducting phase transition as long as the level spacing in a localization volume $\delta$ is much smaller than the superconducting gap, $\Delta$~\cite{ma1985localized}.

\begin{figure*}
    \includegraphics[width = 0.7 \textwidth]{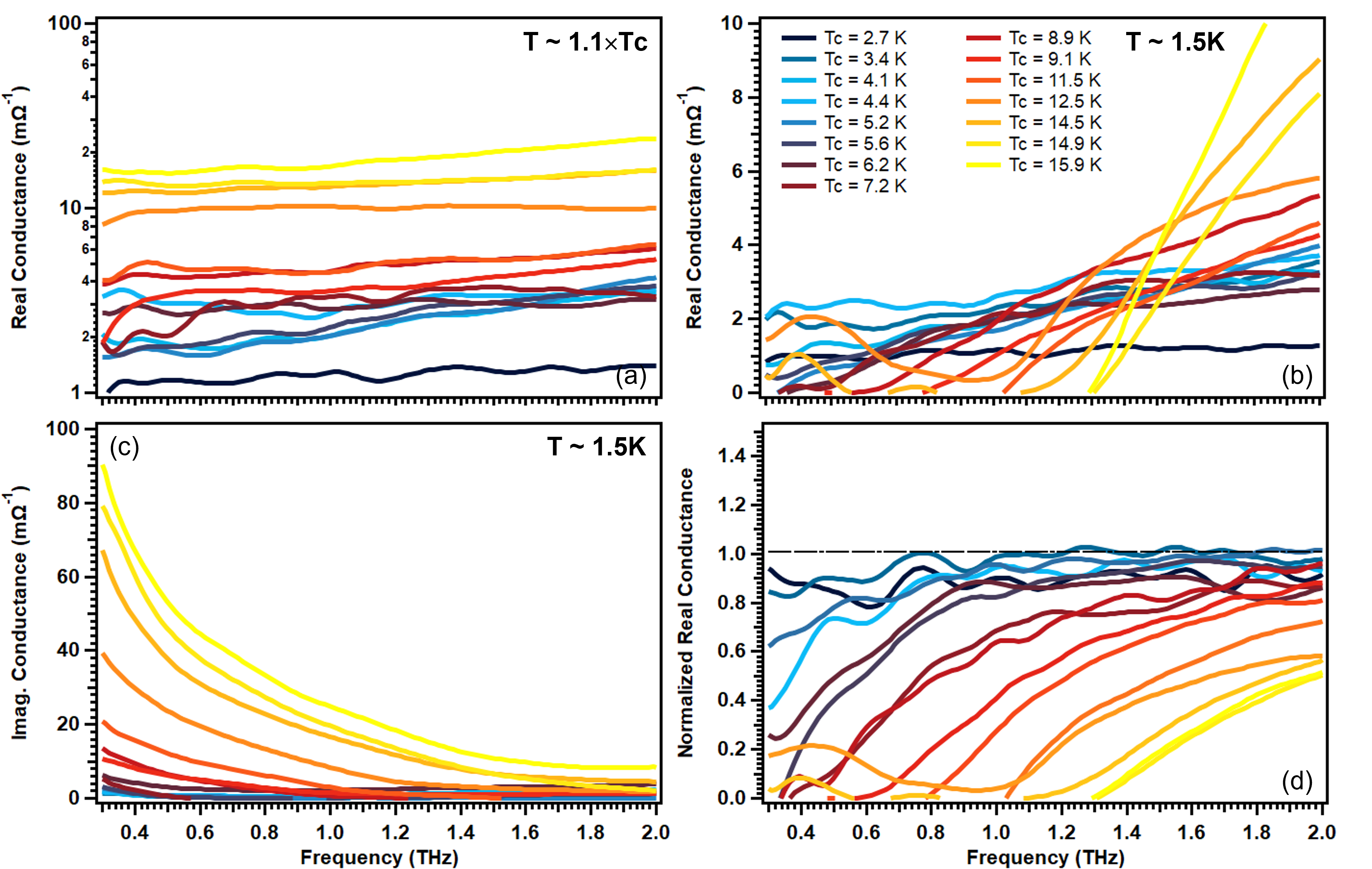}
    \caption{\label{Linear} Real part of the optical conductance of the NbN films in (a) the normal state ($T \approx 1.1\times T_c$) and at (b) $T \approx 1.5 K$ ($T \ll T_c$). (c) The imaginary part of optical conductance at $T \approx 1.5 K$. (d) The real part of the low temperature optical conductance normalized by the normal state optical conductance.  $T_c$ is determined from the temperature where the resistance goes to zero. }
\end{figure*}

While the theory of ``dirty" superconductors is the basis of the understanding of superconductivity in many disordered alloys, poly-crystalline, and amorphous materials, there are serious limitations. These calculations implicitly assume that the pairing mechanism remains unaffected by disorder, which crucially ignores the evolution of electron-electron interactions with disorder. In fact, disorder can amplify repulsion or attraction between electrons depending on details of Coulomb repulsion, which then can suppress or increase $T_c$\cite{finkel1987superconductivity,feigel2007eigenfunction,burmistrov2012enhancement}.  This can be viewed as an effective increase in the ``interaction time" between slowly diffusing electrons.  The theory also needs revision in the regime $l \approx \hbar/p_F$ where disorder driven localization mandates destruction of extended electronic states that can sustain diffusive transport.  It has been argued that if one neglects Coulomb repulsion and critical fluctuations, true long-range ordered superconductivity can be supported on the insulating side of the metal-insulator transition (MIT) as long as the superconducting energy gap remains larger than the separation of energy levels in a localization volume \cite{ma1985localized, kapitulnik1985anderson, kotliar1986anderson}. In addition, a number of experiments suggest that finite superconducting correlations exist even in the disorder or magnetic field driven insulating state.  These include a large magneto-resistance peak \cite{nguyen2009observation, baturina2007quantum}, signatures of local phase coherence via 2e Little-Parks oscillations \cite{stewart2007superconducting}, finite superfluid stiffness in the ac conductivity above $T_c$ \cite{liu2011dynamical,crane2007survival}, and the emergence of a ``pseudogap'', defined as a suppression in the low energy density of states above $T_c$ \cite{sacepe2010pseudogap, mondal2011phase,chand2012phase}.   Observation of large Nernst effect in the normal state of amorphous superconducting films \cite{pourret2006observation} and more recent shot noise measurements in disordered superconducting thin films \cite{doi:10.1126/science.abe3987} also provide strong evidence for the presence of Cooper pairs at temperatures several times greater than $T_c$. Moreover, there is experimental evidence and theoretical support for a scenario where superconducting correlations become extremely inhomogenous at large disorder even in systems that are nominally homogeneous~\cite{ghosal1998role,ghosal2001inhomogeneous,feigel2007eigenfunction,sacepe2011localization,chand2012phase,kamlapure2013emergence}. 

\begin{figure}
    \includegraphics[width = 0.4\textwidth]{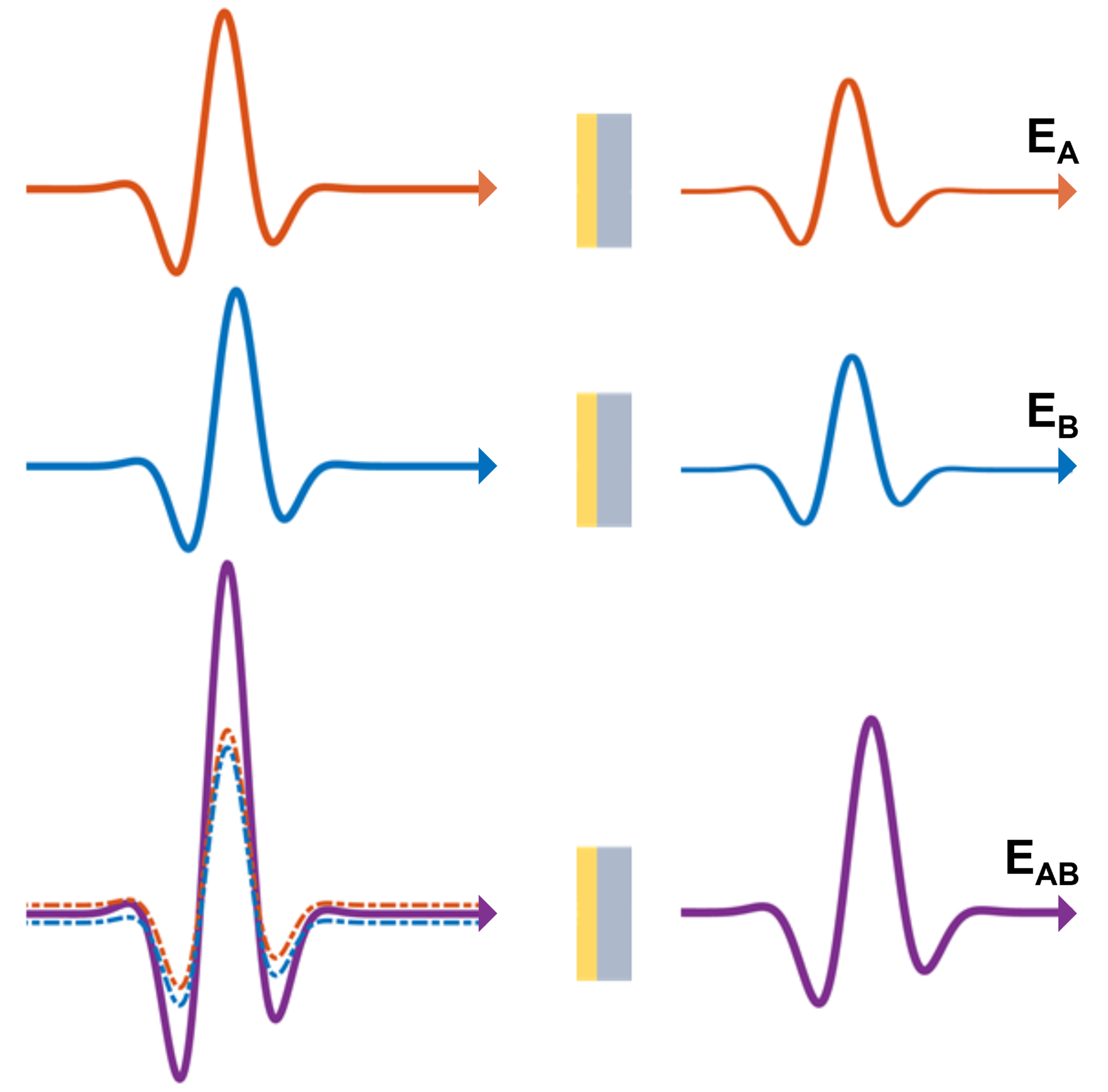}
    \caption{\label{setup} Schematic of the experimental protocol. The transmitted THz field in response to two intense THz pulses $E_A$ and  $E_B$ are measured independently and in conjunction $E_{AB}$. The nonlinear response is then quantified as $E_{NL} = E_{AB} - E_{A} - E_{B}$.}
\end{figure}

\begin{figure*}
    \includegraphics[width = \textwidth]{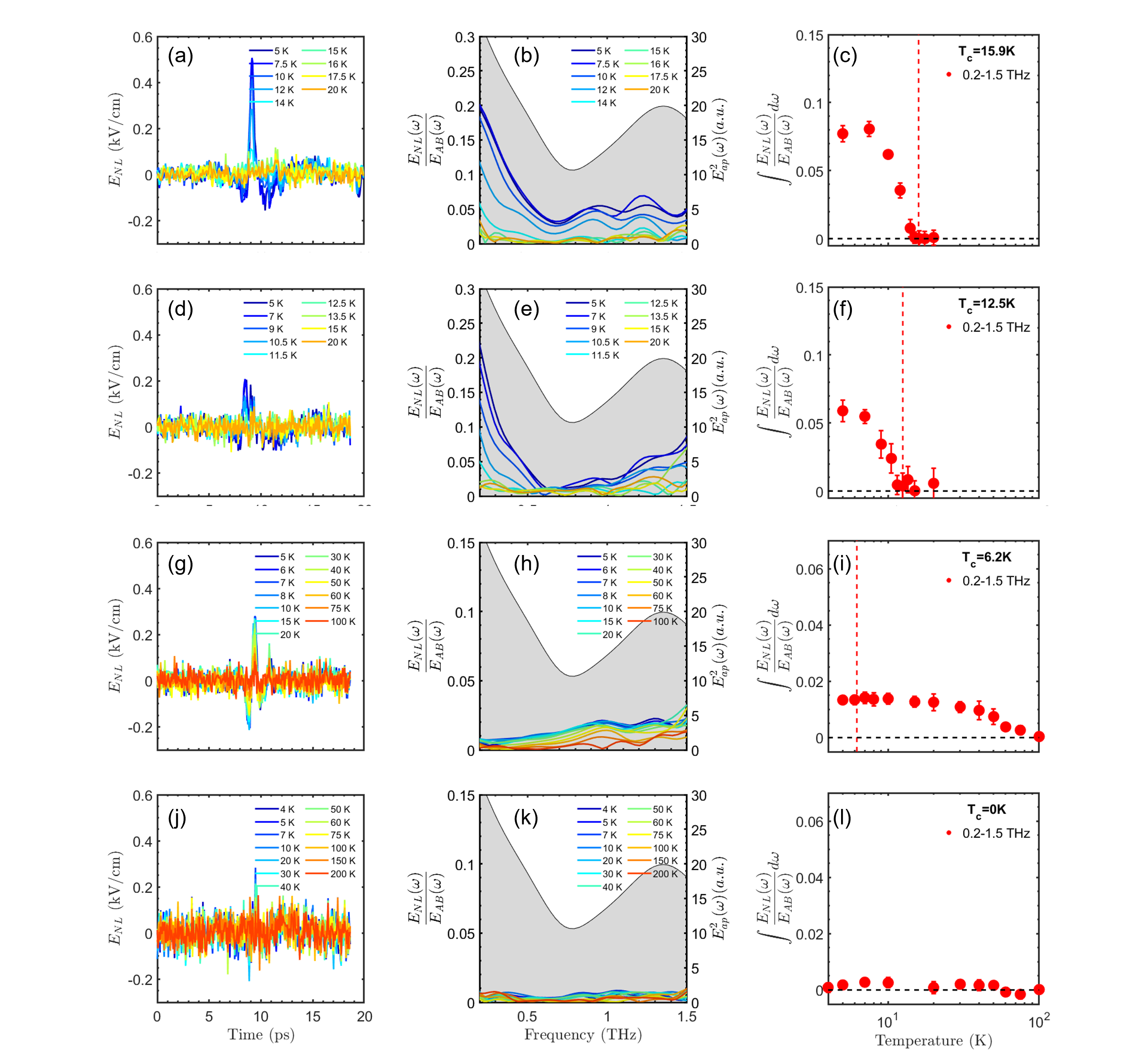}
    \caption{\label{ENL} nonlinear THz response for four different disorder level with $T_c$ 15.9K (a-c), 12.5K (d-f), 6.2K (g-i) and insulating sample (j-l). The first (a,d,g,j) column shows the nonlinear response as measured in time-domain. The second column, (b,e,h,k) shows the ratio of the nonlinear spectra normalized to $E_{AB}$. The grey shaded region is the area under the square of the applied electric field. Third column (c,f,i,l) shows the normalized nonlinear spectrum as shown in the second column integrated over the frequency range 0.2-1.5 THz.   The red dashed line shows $T_c$. A small background has been subtracted which to account for the integral of noise floor within the spectral region.}
\end{figure*}

\begin{figure}
    \includegraphics[width = 0.5\textwidth]{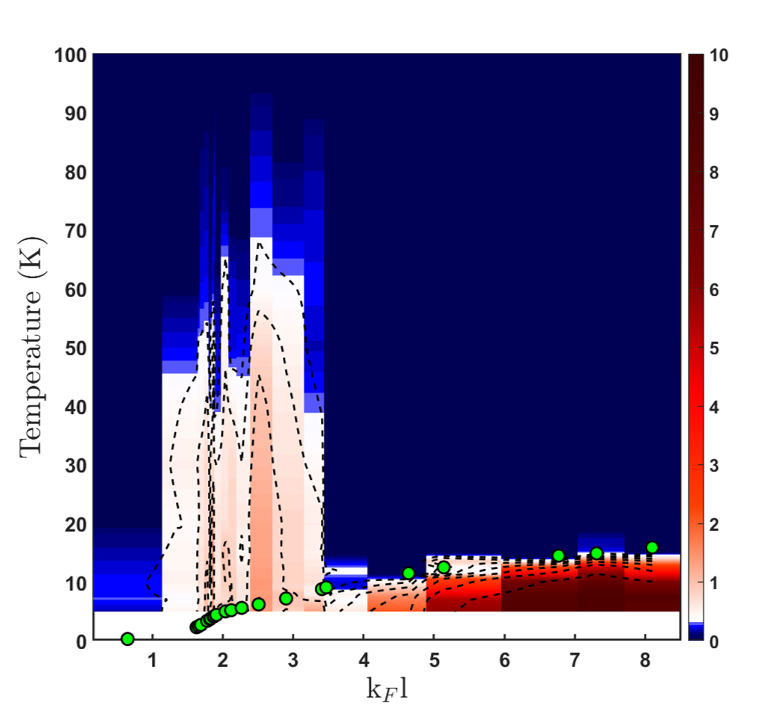}
    \caption{\label{phase} Color map showing the frequency integrated normalized nonlinear response for all NbN samples ($\times 100$) as a function of temperature and $k_Fl$.   Green dots are the co-plotted $T_c$ of the samples.}
    \label{PhaseDiagram}
\end{figure}

Low energy THz-range nonlinear spectroscopy has emerged as a new tool for probing superconductivity and other quantum properties and have lead to insights that were hitherto inaccessible with other tools. Recent experiments have demonstrated that nonlinear response can be used as a sensitive probe of symmetry breaking in solids, Berry-phase effects in topological materials \cite{orenstein2021topology, ma2021topology} and has offered unique access to nonequilibrium phases in driven systems \cite{mankowsky2016non}. Higher harmonic THz generation and optical Kerr effect in THz field driven superconductors has been interpreted as a direct probe of the amplitude mode of the superconducting order parameter in conventional and unconventional superconductors \cite{shimano2020higgs}.

In Fig. \ref{Linear}(a) we show the real part of the linear response optical conductance of NbN films with different $T_c$ in the normal state at $T \approx 1.1\times T_c$.  In accord with previous studies on disordered NbN~\cite{cheng2016anomalous}, the most disordered samples show increasing signatures of localization, with the dissipative conductivity becoming a strongly increasing function of frequency for the most disordered samples.  As films are cooled below the critical temperature, spectral weight at energies below the superconducting gap $2\Delta$ is transferred to the Dirac-$\delta$ peak at zero frequency as shown in Fig. \ref{Linear}(b).  The $2\Delta$ gap is progressively suppressed with increasing disorder.  Signatures of the formation of the superconducting condensate can also be observed in the imaginary part of the optical conductance at low temperature as shown in Fig. \ref{Linear}(c) which diverges as $\sim 1/\omega$ as $\omega \rightarrow 0$ with a pre-factor proportional to the superfluid density (e.g. the spectral weight in the $\delta$ function). The spectral weight transfer is perhaps most apparent in Fig. \ref{Linear}(d) when the real part of optical conductance at the base temperature below $T_c$ is normalized to the normal state conductance. As disorder is increased, $T_c$ is reduced by an order of magnitude and corresponding lowering of $2\Delta$ and superfluid density is observed in the linear response.  All these results are in excellent agreement with previous THz studies on disordered NbN films~\cite{cheng2016anomalous}.

To explore the nonlinear response, intense THz pulses were generated by exciting a LiNbO$_3$ crystal with amplified 800 nm laser pulses with $\approx$3 mJ energies in the tilted-pulse-front scheme \cite{hebling2002velocity, hirori2011single} and detected with standard electro-optic methods. The peak electric field strength of the THz pulses is $\approx$25 kV/cm. We employ a two pulse measurement scheme to directly quantify the nonlinear response. We measure the transmitted light through the sample in response to two intense THz pulses, $A$ and $B$ and measure $E_A(t)$, $E_B(t)$ and $E_A(t)+E_B(t)$ ($E_{AB}(t)$) independently. The nonlinear response is then directly obtained as $E_{NL}(t) = E_{AB}(t) - E_{A}(t) - E_{B}(t)$. A schematic of the experiment is given in Fig. \ref{setup} and more details can be found in ~\cite{mahmood2021observation}.

Quite generally, strong nonlinear THz electrodynamics is expected in a conventional superconductor.  This can be seen in a number of ways.  At the Ginzburg-Landau level  the superfluid density (in a dirty limit superconductor) is function of the superfluid velocity $v$, which gives a contribution to the current that goes like $v^3$.  In phase only $x-y$ models, the expansion of the cosine term gives a similar nonlinearity that is also proportional to the superfluid density.  More refined treatments show that the nonlinear coupling of light may excite charge density fluctuation at $2\Delta$, amplitude modes of the superconducting order parameter, and manifest via the nonlinear Meissner effect \cite{matsunaga2014light,shimano2020higgs,cea2016nonlinear,yip1992nonlinear}.  In contrast, a normal metal with a parabolic band and only elastic scattering has an electrodynamic response that is strictly linear \cite{wolff1966theory,rustagi1969effect,rustagi1970effect,yuen1982difference}.

We investigated the THz nonlinearity as function of disorder in samples with $T_c$ of $\approx$16K to 0K.  Fig \ref{ENL} shows the raw nonlinear response in the time domain for a number of samples spanning this disorder range.  The leading order nonlinearity in centrosymmetric NbN is a third-order $\chi^{(3)}$ response~\cite{matsunaga2014light}. Therefore the plotted signals can be taken as approximately proportional to $\chi^{(3)}$. The measured nonlinear spectra $E_{NL}(\omega)$ was normalized by $E_{AB}(\omega)$ to account for the the linear response. The grey shaded area in the middle column is representative of the Fourier transform of the square of the incident electric field, $E^2_{ap}(t)$.

For the lowest disorder (highest-$T_c$) samples, the nonlinear spectra (Fig. 3(b)) shows a maximum at the lowest energy.  One expects a resonant feature when the photon energy is equal to 2$\Delta$ ~\cite{cea2016nonlinear,udina2019theory} but this is hard to isolate in the current experimental scheme with the broadband THz pulse.  The temperature dependence of the nonlinearity (Fig. 3(c)) tracks the superconducting response in that it decreases as one approaches $T_c$ from below and disappears above the superconducting transition.  This is behavior that looks much like the behavior of the superfluid density in accord with simple expectations. With increasing disorder, we observe an expected reduction in the overall magnitude of the nonlinear response but its temperature dependence remains strictly correlated with the superconductivity as shown in Fig. \ref{ENL}(d-f). Based on transport and tunneling experiments, this is roughly the same as the regime of ``dirty'' superconductivity where a suppression of $T_c$ is observed but the overall phenomenology is still BCS-like with a reduced $\Delta$ that is proportional to $ k_B T_c$.

As $T_c$ is further suppressed by a little over twice the clean limit, we observe a qualitative change in the THz nonlinearity. While the overall magnitude remains lower than that in the clean samples, the temperature dependence does not track the superconductivity according to the conventional BCS estimate of an order parameter. The response remains nearly constant far beyond $T_c$ and only decays to zero at a much higher temperature scale the maximum of which is $\approx$ 60K ($\sim 4\times T_c$ of the cleanest NbN) as shown in Fig. \ref{ENL} (g-i). While tunneling experiments in this range observe a pseudogap phase above critical temperature~\cite{chand2012phase}, it must be noted that our THz nonlinearity persists far above the onset temperature of the pseudogap, which is approximately 7K for highly disordered NbN samples near the MIT.  At even higher disorder levels, $T_c$ goes to zero as superconductivity is suppressed altogether, no sign of superconductivity is observed in linear response, and the THz nonlinearity disappears as well.

Fig. \ref{phase} is a color plot of the the integrated nonlinearity for the entire range of disorder levels starting on the right with highest $T_c$ all the way into the insulating regime when $T_c$ goes to zero. Again, one can see for the regime of least disordered samples, the extent of the THz nonlinearity tracks the superconducting state itself.  However, as $T_c$ falls below $\sim$ 7.5 K, a ``bubble" of highly anomalous nonlinear THz response is observed.   For the sample with $T_c  = 6.2$K, this response is observed to as high as 60K, which is approximately 10 times $T_c$. The disorder level in each sample has been quantified using the Ioffe-Regel parameter $k_Fl$, and estimated based on previous experiments on similar NbN films \cite{chand2012transport}.

What is the origin of this remarkable nonlinearity?   Both anomalous aspects of the highly disordered normal state conductor and vestiges of superconductivity to high temperature must be considered as possibilities.   As mentioned above, normal metals with parabolic bands (in which velocity changes are proportional to changes in momentum) and which are dominated by elastic scattering have zero nonlinear response~\cite{wolff1966theory,rustagi1969effect,rustagi1970effect,yuen1982difference}.   Metals with strong inelastic scattering can have large nonlinear effects, but it seems unlikely that such effects dominate NbN near the disorder driven metal-insulator transition.   nonparabolicity of bands in NbN could cause nonlinearity, but very small nonlinearity is observed in the normal state of the least disordered samples and one expects that any nonlinearity decreases with increased disorder (see Supplementary Material \cite{SI}).  At finite frequency, a modest enhancement of the nonlinearity could occur if the energy relaxation rate of driven carriers is larger than the momentum relaxation rate\cite{yuen1982difference}, but this seems an unlikely scenario near a disorder driven metal-insulator transition.  We see no reasonable mechanism that the nonlinear response could derive from some enhanced nonlinearity of the extremely disordered normal state.  Please see the SM for further discussion.

The fact that the $\chi^{(3)}$ signal appears to fall to zero when $T_c$ is suppressed to zero, and the fact that the response is smooth through the transition strongly suggests that despite its remarkable persistence to very high temperatures, the nonlinearity has a superconducting origin.  It is well established that the $T_c$ of some superconductors with low superfluid density (decreased by disorder or correlations) is set not by the energy scale of Cooper pair formation, but by the superfluid stiffness which is the energy scale for disordering the phase stiffness of the superconducting order parameter~\cite{emery1995superconductivity}.  Such systems -- for instance highly disordered superconductors near the MIT -- can show remnant signatures of Cooper pair formations to temperatures well higher than $T_c$.  For instance, both amorphous InO$_x$ and NbN show tunneling gaps that for very disordered samples persist to temperatures well above $T_c$~\cite{sacepe2011localization,chand2012phase}.  In NbN, at low disorder ($k_Fl > 4 $), the tunneling gap follows the mean field Bardeen-Cooper-Schrieffer behavior and the superconducting energy gap vanishes at the same temperature where electrical resistance becomes finite.  For stronger disorder ( $k_Fl < 4$ ) a ``pseudogap” state emerges where a tunneling gap persists up to temperatures as high as 7K.   Optical conductivity experiments on disordered InO$_x$ and cuprate superconductors reveal finite frequency superconducting fluctuations in a regime not in excess of 25$\%$ above $T_c$~\cite{corson1999vanishing,bilbro2011temporal,liu2011dynamical}.  Such results suggest that Cooper pairs continue to exist in the system even after the zero resistance state is destroyed.  However, in all cases the onset temperature of this ``pseudogap" is well below the $T_c$ of the cleanest samples.   This is in obvious contrast with the present case where the $\chi^{(3)}$ signal persists to more than 4 times the $T_c$ of the cleanest sample.

As discussed above, a remarkable prediction of theories of superconducting systems near a localization transition is an enhancement of the superconducting transition temperature, and/or an increase of the gap energy~\cite{ghosal2001inhomogeneous, feigel2007eigenfunction,burmistrov2012enhancement}.  These features can be seen as a consequence of an enhanced ``interaction time" between slowly diffusing electrons.  However, as seen in Fig.~\ref{PhaseDiagram}, $T_c$ does {\it not} increase close to metal-insulator transition.   Nevertheless, we would propose that much in the same fashion that finite frequency optical experiments can reveal fluctuating superconducting correlations ~\cite{corson1999vanishing,bilbro2011temporal,liu2011dynamical}, the finite frequency nonlinear THz experiments can reveal fluctuating superconductivity even to regions where superconducting long-range order does not occur.  We would also point out that models for multi-fractal superconductivity~\cite{feigel2010fractal} give a Ginzburg number of order unity, showing that such models are expected to exhibit a very large fluctuation range.   In probing order parameter fluctuations, the present experiments have the advantage over the linear optical response that these are essentially background free as the normal metallic state has vanishing $\chi^{(3)}$ nonlinearity.   Going forward, it will be important to have explicit theoretical calculations of the  $\chi^{(3)}$ in the fluctuation range of the superconducting response.  One may also speculate how one might stabilize this superconducting correlations to be true long-range ordered superconductivity at these unprecedentedly large temperatures for ``conventional superconductors".

\bigskip

\section*{Materials and Methods} 
Thin epitaxial films of NbN were deposited on (100) oriented MgO substrate using reactive dc magnetron sputtering using a Nb target. The thickness of these films are $\approx$20nm which is thicker than the 4-8 nm coherence length in these systems~\cite{mondal2011phase}.  The disorder is tuned by controlling the level of Nb vacancies in the lattice, which is controlled by the Ar:N ratio in the sputtering chamber.  Further details of the sample preparation can be found in Ref.~\cite{chockalingam2008superconducting}.

Nonlinear THz spectroscopy was performed using two intense THz pulses (A and B) generated by the tilted pulse front technique such that the pulses are overlapping in time. The two THz pulses are focused onto the sample in a collinear geometry \cite{mahmood2021observation}. The transmitted THz fields were detected by standard electro-optic (EO) sampling using a 0.5 mm GaP crystal. Displayed data was taken with a maximum electric field being \SI{25}{kV/cm}  on the sample for each pulse.  A differential chopping scheme is used to extract the nonlinear signal ($E_{NL}(t) = E_{AB}(t) - E_{A}(t) - E_{B}(t)$) resulting from the interaction of the two THz pulses with the sample. $E_{AB}$ is here refers to the transmitted signal when both THz pulses are present while $E_{A}$ and $E_{B}$ are the transmitted signals with each pulse A and B present individually. The THz linear optical response of these samples has been performed using custom built Auston switch based time-domain THz spectroscopy. Details of the experimental setup is given in ref. \citep{laurita2016modified}. The cryogenic cooling of the sample for both the linear and nonlinear spectroscopy was done using a continuous flow helium cryostat. 

\bigskip

\section*{Data availability}
All data that support the plots within this paper and other findings of this study are available from the corresponding author upon reasonable request

\bigskip

\section*{Acknowledgements}
We would like to thank L. Benfatto, M. Feigel'man, K.C. Rustagi, N. Trivedi for a helpful comments and conversations.  This project was supported by a now canceled DARPA DRINQS program grant and by the Gordon and Betty Moore Foundation, EPiQS initiative, Grant No. GBMF-9454.

\bigskip

\section*{Author contributions}
DC, DB and FM built the nonlinear THz setup and performed experiments.  DC analyzed the data.  RR performed the THz linear response measurements.  JL performed dc resistivity experiments.  JJ and PR grew and characterized the films using mutual inductance technique.   NPA directed the project.   All authors contributed to the writing and editing of the manuscript.

\bigskip

\section*{Competing interests}
The authors declare no competing interests.

\bigskip
\balancecolsandclearpage
\onecolumngrid
\section*{Supplementary Materials:  Anomalous high-temperature THz non-linearity in superconductors near the metal-insulator transition}

\twocolumngrid
\section{THz Nonlinearities of a degenerate Fermi gas}

\begin{figure*}
    \includegraphics[width = \textwidth]{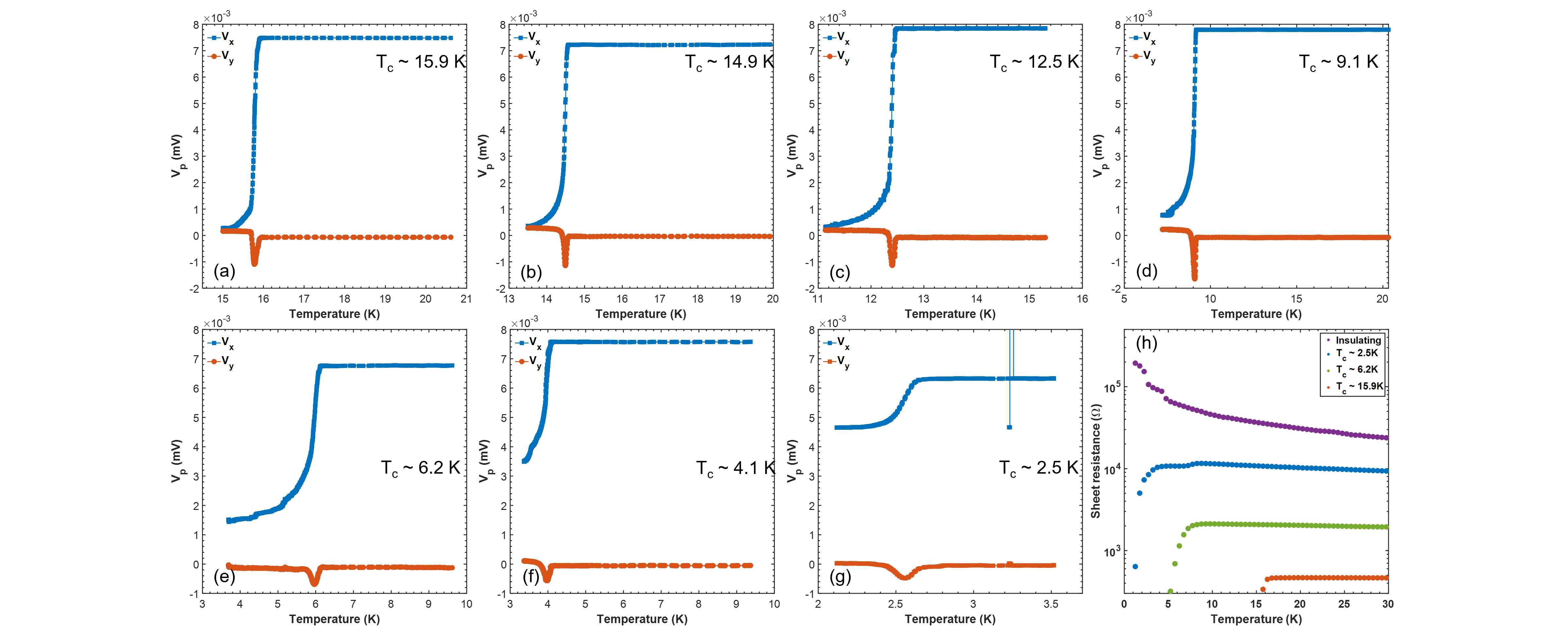}
    \caption{\label{transport} (a)-(g) The real (blue) and imaginary (red) part of the mutual inductance measured as a function of temperature demonstrating the superconducting transition. (h) Sheet resistance of a few representative samples measured as a function of temperature. Note that no signature of a superconducting T$_c$ has been observed for the insulating sample.}
    \label{transport}
\end{figure*}

The subject of THz electrodynamic nonlinearities for degenerate Fermi systems was investigated extensively in the 60s and 70s in the context of in narrow gap semiconductors~\cite{wolff1966theory,rustagi1969effect,rustagi1970effect}.   It was found that appreciable nonlinearities can arise in systems with nonparabolic bands and/or systems with strongly energy dependent scattering.   We expect that the former is much more dominant than the latter in disordered systems near the disorder driven metal-insulator transition.

Non-parabolicity as a source for THz nonlinearity was first investigated by Wolff and Pearson~\cite{wolff1966theory} to explain the large nonlinearities in doped InAs and InSb discovered with Q-switched CO$_2$ lasers\cite{patel1966optical}.   Among the work that followed was that of Rustagi who solved the Boltzmann equation for the motion of electrons in a spherically symmetric system by looking at driven changes in the distribution function $f(k, t)$ expanded in spherical harmonics $Y_{lm}$. A perturbation $f(k, t) Y_{lm}$ was assumed to decay exponentially at a rate $\gamma_l$ with a Raman-like energy transfer between electric fields at frequencies $\omega_1$ and $\omega_2$ ~\cite{rustagi1969effect,rustagi1970effect}.  Later, Yuen and Wolff presented a scenario in terms of the kinetic equations for momentum and energy relaxation~\cite{rustagi1969effect,rustagi1970effect}.  Although they used different calculational schemes, the resulting equations were largely similar to Rustagi with relaxation rates that were explicitly the rates of energy and momentum dependent relaxation.   Below, we use the language of Yuen and Wolff as they have more directly physical intuitive meaning.  In either case, the nonlinearities can be understood heuristically as arising from the effective energy dependent mass that the charges experience as they are ``sloshed" by strong electric fields in a slightly non-parabolic band.  With an electronic dispersion of an inversion symmetric system that can be approximated as 
$E(k) = \frac{\hbar^2 k^2}{2 m_*} - \beta k^4 $, the dominant contribution to the nonlinear 3rd order response is

\begin{equation} \label{eq1}
\begin{split}
\chi^{(3)}(\omega_3 = 2\omega_1 -\omega_2) & = \\  \frac{1  }{2 \omega_1 - \omega_2 } &  \frac{-n e^4 \beta   } {  (\omega_1 - i\gamma_m)^2(\omega_2 + i\gamma_m)  }  \\  & \; \; \; \; \; \; \Big( 1 + \frac{ 2 \gamma_m - \gamma_e }{\gamma_e - i (\omega_1 - \omega_2 ) } \Big).
\end{split}
\end{equation}

Here n is the electron density, $m_*$ is the band mass at the bottom of the band, and $\beta$ is the quartic correction to the dispersion.  $\gamma_e$ and $\gamma_m$ are the energy and momentum relaxation rates.   Deviations from parabolicity not captured by the $\beta$ term only manifest as a modified prefactor of this expression.  Although similar dependencies on at least two different relaxation rates are found in the expression derived by Rustagi~\cite{rustagi1969effect,rustagi1970effect}, this one also has a Drude-like prefactor that causes the response to be strongly suppressed with large momentum (elastic) scattering. One can regard this as an expression analogous to an effective Drude-like model for quasi-free electrons, but for the 3rd-order nonlinear response.

For large elastic scattering e.g. $  \gamma_m \gg \omega_1, \omega_2, \omega_3, \gamma_e $ this expression simplifies to

\begin{equation} \label{eq2}
\begin{split}
\chi^{(3)}(\omega_3 = 2\omega_1 -\omega_2)  = & \\ \; \; \; \; \frac{1  }{2 \omega_1 - \omega_2 }   \frac{-2i n e^4 \beta  }{ \gamma_m^2  } &  \Big( \frac{ 1 }{\gamma_e - i (\omega_1 - \omega_2 ) } \Big).
\end{split}
\end{equation}

As expected the $\chi^{(3)}$ that arise from non-parabolicity will be  strongly suppressed by strong disorder scattering with large $\gamma_m$.

As discussed in the main text, the electrodynamic nonlinearities for a classical plasma with a parabolic dispersion are extremely small as they are driven in part by the light's oscillating magnetic field through a ${\bf v} \times {\bf B} $ effect~\cite{bloembergen1966optical}.  Two optical fields can mix to excite electron density oscillations at $\omega_2 \pm \omega_1$.  This effect can be regarded semiclassically for co-polarized $z$ directed waves as a transverse $E_x$ field at $\omega_1$ driving a current that is then acted upon by an $B_y$ field at $\omega_2$, which induces density fluctuations modulated in the $z$ direction.   In a bulk material, these longitudinally polarized density fluctuations cannot themselves radiate.  However they can Raman scatter a third photon of frequency $\omega_1$.  The effect is largest on resonance when $\omega_p = \omega_2 \pm \omega_1$ can excite a longitudinal plasma mode.   The effect goes like $1/\epsilon( \omega_1 \pm \omega_2 ) $ and so it is expected that the nonparabolic effect discussed above is larger by a factor of $ 4 m^3 c^2 \beta \epsilon( \omega_1 \pm \omega_2 )$.  Even on-resonance the dielectric constant $\epsilon( \omega_p  ) $ is $ \frac{-i}{\omega \tau}$ gives an effect that is estimated to be smaller than the non-parabolic one for small gap semiconductors by a factor of at least $\sim$100~\cite{wolff1966theory}.  In the off-resonant range of $\omega \ll \omega_p$ that is relevant to our experiment, one expects that this effect is even much more strongly suppressed as charge density fluctuations are screened by the almost divergent metallic dielectric constant $\epsilon(\omega \rightarrow 0)$, and is accordingly vanishingly small. 

\section{Transport measurements}
Nominal T$_c$ for the samples were determined using two-coil mutual inductance measurements. The films were coaxially sandwiched between a quadrupole primary coil and a dipole secondary coil. Mutual inductance between the primary and  the secondary coil was measured as a function of temperature by passing a small ac current through the primary coil and measuring the induced voltage, $V_p$ at secondary pick-up coil using lock-in techniques. The in-phase and out-of-phase components of the pick-up voltage $V_p$ and is related to the real and imaginary part of the mutual inductance. One expects a dip in the real part and a peak in the imaginary part at the critical temperature which correspond to the inductive and resistive coupling between the two coils respectively. Details on the experimental setup and technique can be found in Ref. \cite{kamlapure2010measurement}.

Fig. \ref{transport} (a)-(g) shows the in-phase and out-of-phase components of the measured voltage for some of the samples used in this study. Clear signature for critical temperature was observed for all samples except for the insulating sample, discussed in the main text. Sheet resistance measured as a function of temperature shown in Fig. \ref{transport} (h) are also in good agreement with the mutual inductance data. Signature of superconducting transition for all samples except for the insulating one.

\end{document}